\documentclass[reprint,superscriptaddress,amssymb,amsmath,aps,showpacs,10pt,floatfix,longbibliography,pra]{revtex4-1}
\pdfoutput=1

\usepackage{color}
\usepackage[colorlinks,bookmarks=false,citecolor=darkblue,linkcolor=red,urlcolor=blue]{hyperref}
\def\cred{}
\definecolor{darkred}{rgb}{0.7,0.0,0.0}

\definecolor{darkblue}{rgb}{0,0.02,0.45}

\definecolor{darkgreen}{rgb}{0.02,0.45,0.0}

\definecolor{violet}{rgb}{0.8,0.2,0.6}

\usepackage[pdftex]{graphicx} 
\usepackage{epstopdf}
\usepackage{verbatim}
\usepackage{amsmath}
\usepackage{color}
\usepackage{subfigure}
\usepackage{amsbsy}
\usepackage{wasysym}

\begin{document}

\title{Spin liquids in geometrically perfect triangular antiferromagnets}

\author{Yuesheng Li}
\affiliation{Experimental Physics VI, Center for Electronic Correlations and Magnetism, University of Augsburg, 86159 Augsburg, Germany}
\affiliation{Wuhan National High Magnetic Field Center and School of Physics, Huazhong University of Science and Technology, 430074 Wuhan, China}

\author{Philipp Gegenwart}
\affiliation{Experimental Physics VI, Center for Electronic Correlations and Magnetism, University of Augsburg, 86159 Augsburg, Germany}

\author{Alexander A. Tsirlin}
\email{altsirlin@gmail.com}
\affiliation{Experimental Physics VI, Center for Electronic Correlations and Magnetism, University of Augsburg, 86159 Augsburg, Germany}

\begin{abstract}
The cradle of quantum spin liquids, triangular antiferromagnets show strong proclivity to magnetic order and require deliberate tuning to stabilize a spin-liquid state. In this brief review, we juxtapose recent theoretical developments that trace the parameter regime of the spin-liquid phase, with experimental results for Co-based and Yb-based triangular antiferromagnets. Unconventional spin dynamics arising from both ordered and disordered ground states is discussed, and the notion of a geometrically perfect triangular system is scrutinized to demonstrate non-trivial imperfections that may assist magnetic frustration in stabilizing dynamic spin states with peculiar excitations.
\end{abstract}

\maketitle

\section{Introduction}
Frustrated magnets entail competing exchange interactions and behave differently from their non-frustrated counterparts, where all couplings act to stabilize a magnetic order. It is then natural that in the presence of frustration any ordering effects are impeded, and a non-ordered, paramagnetic-like state survives down to much lower temperatures than in conventional magnets. Eventually, frustrated spin systems may not show any long-range magnetic order at all, and enter instead a peculiar low-temperature state known as \textit{spin liquid}. On the most basic level, this state can be matched with ordinary liquids in the sense that spins develop short-range correlations but lack any long-range magnetic order, and the system does not undergo symmetry breaking upon cooling.

An exact definition of a spin liquid is, however, more complex than that and involves a fair amount of ambiguity or even contention. Key ingredients of the spin-liquid state are the absence of long-range magnetic order and the presence of persistent spin dynamics down to zero temperature. These two aspects essentially distinguish the spin liquid from magnetically ordered states with symmetry breaking, and from spin glasses, where spin fluctuations slow down upon cooling, so that spins eventually become static. On the other hand, without postulating any microscopic aspects of this disordered and dynamic (liquid-like) state, such a definition allows fundamentally different systems with similar phenomenology to be classified as spin liquids.

Microscopically, one distinguishes between quantum spin liquids where spins are quantum-mechanically entangled, and classical spin liquids that can be naively seen as a ``soup'' of different magnetic orders, all having the same energy. This classical scenario leaves spins to fluctuate as long as the temperature is high enough to overcome transition barriers between different ordered states. At very low temperatures, classical spin liquid is expected to freeze, whereas its quantum counterpart remains dynamic by virtue of quantum fluctuations. Experimentally, this presence or absence of spin freezing serves as a useful diagnostic tool along with magnetic excitations that bear signatures of many-body entanglement in quantum spin liquids although may be exotic in classical spin liquids too~\cite{balents2010,gingras2014}.

The gap between these two definitions -- phenomenological and microscopic -- has led, and still leads to a common confusion of whether a given material should be interpreted as a spin liquid. If it is, how to juxtapose the experimental magnetic response with theory, and if it is not, is there still a room for the spin-liquid physics? It is exactly these controversial but pertinent issues that we seek to address in the present brief review using triangular antiferromagnets as an example. For a more detailed introduction into the physics of spin liquids beyond the triangular systems we refer readers to excellent summaries~\cite{knolle2019,broholm2019}, as well as more technical and elaborate overviews of the field~\cite{savary2017,zhou2017} that were recently published.

Historically, triangular antiferromagnets were first systems where magnetic frustration was encountered, and early ideas of quantum spin liquid state were established~\cite{anderson1973}. On the experimental side, the field has seen several revivals related to active studies of organic charge-transfer salts and, more recently, Co$^{2+}$ and Yb$^{3+}$ oxide compounds that are the main topic of our present review. For practical purposes, we restrict this review to geometrically perfect spin-$\frac12$ triangular antiferromagnets, where magnetic ions form regular triangular framework, and exclude systems with higher spin as well as systems with three-fold frustrated loops comprising non-equivalent exchange pathways (an excellent overview of all those can be found in Ref.~\onlinecite{starykh2015}). In particular, organic charge-transfer salts that are often discussed in the context of spin-liquid physics~\cite{powell2011} are beyond the scope of our present review, because their spin lattices entail distorted triangles, and low-temperature structural instabilities abound. 

\begin{figure*}
\includegraphics{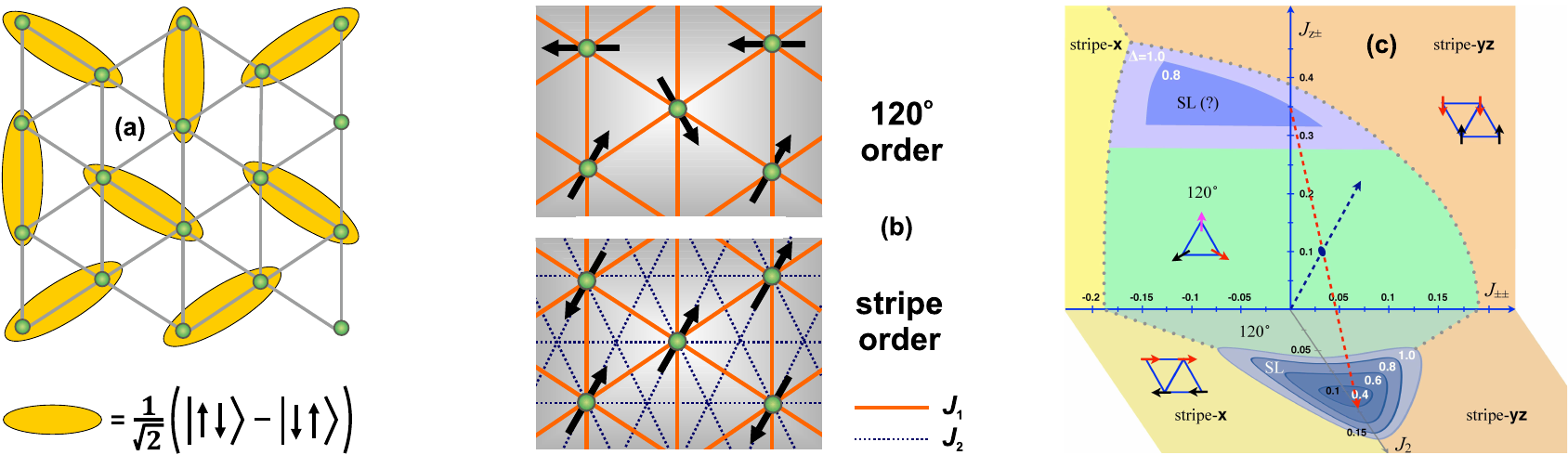}
\caption{\label{fig:triangular}
Spin states in triangular antiferromagnets. (a) Valence-bond solid with valence bonds between nearest neighbors; a superposition of all such states makes the nearest-neighbor RVB state. (b) $120^{\circ}$ and stripe orders. (c) Magnetic phase diagram~\cite{zhu2018} with lines corresponding to different $\Delta$ values in Eq.~\eqref{eq:delta}; SL stands for the spin liquid. Panel (c) is reprinted with permission from Ref.~\onlinecite{zhu2018}, \copyright\  American Physical Society, 2018.
}
\end{figure*}

As simple and natural as it seems, the definition of the geometrically perfect material also appears to be controversial -- perhaps even more controversial than the definition of the spin liquid itself. The natural development of the field requires that every new spin-liquid candidate is claimed to be more perfect and ideal than its predecessors, which raises a question of whether any material is truly ``geometrically perfect''. We shall discuss this issue at some length to show non-trivial departures from the ``perfection'', and argue that they may strongly influence the physics, sometimes in a very interesting way.

\section{Theory}
\subsection{Resonating valence bonds}
\label{sec:phenomenology}
Early ideas of the quantum spin liquid appeared in the context of Heisenberg spins on the triangular lattice. Anderson and Fazekas~\cite{anderson1973,fazekas1974} conjectured that long-range-ordered $120^{\circ}$ N\'eel state (Fig.~\ref{fig:triangular}b) obtained by classical minimization is not the lowest-energy state of a quantum system, where spins gain additional energy by forming pairs, quantum-mechanical singlets. These singlet pairs were termed \textit{valence bonds} and treated using formalism initially developed by Rumer for molecules~\cite{rumer1932a,rumer1932b,pauling1933a}. The intuitive chemical analogy was further backed by the fact that valence bonds not only occur as spin-pairs but can also form resonant configurations akin to Pauling's idea of resonating bonds in molecules~\cite{pauling1933b}. The presence or absence of the resonance effect distinguishes valence-bond solid (VBS), the combination of static and non-resonating valence bonds on a lattice, from the resonating-valence-bond (RVB) states obtained by a superposition of different VBS states. It is this RVB state with singlet pairs restricted to nearest neighbors (Fig.~\ref{fig:triangular}a), that was proposed as the lowest-energy, liquid-like ground state of triangular antiferromagnets~\cite{anderson1973,fazekas1974}.

Both VBS and RVB states would fall under the most general definition of a spin liquid in the sense that they lack long-range magnetic order and show persistent spin dynamics. However, the VBS state involves in fact some symmetry breaking, although it now relates to singlet pairs and not to individual spins. In real materials, this order becomes even more tangible, because VBS states are intertwined with lattice distortions that stabilize singlet pairs on bonds with stronger exchange interactions~\cite{powell2011}. 

\begin{figure*}
\includegraphics{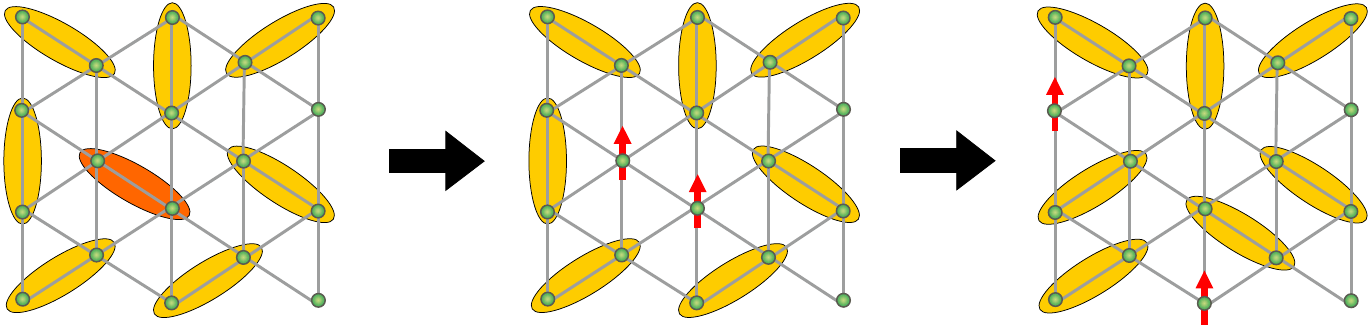}
\caption{\label{fig:spinon}
Spinon excitation in the nearest-neighbor RVB state of triangular antiferromagnets. The excitation is created by breaking one of the valence bonds. Two unpaired spins can propagate independently and constitute spin-$\frac12$ (spinon) excitations.
}
\end{figure*}

Another, and more crucial difference is that only the RVB states show fractional, fermionic \mbox{(spin-$\frac12$)} statistics of magnetic excitations that can be contrasted with the bosonic (spin-1) statistics of magnons in long-range-ordered (anti)ferromagnets. {\cred Excitation brings spin dimer into a triplet state, which can be thought as a superposition of classical states with two parallel spins.} In the RVB states, these spins can separate from each other and move through the crystal independently, because the ground-state wavefunction embraces all possible configurations of valence bonds. In this case, the spin-1 excitations break down (fractionalize) into two \textit{spinons} (Fig.~\ref{fig:spinon}) that become elementary excitations of an RVB state~\cite{kivelson1987,anderson1987}. Emergent spinon excitations reflect the highly non-trivial, entangled nature of RVB quantum spin liquids and bear direct relation to charge fractionalization of fractional quantum Hall state in electronic systems with quantum entanglement~\cite{kalmeyer1987,laughlin1988,kalmeyer1989}. Topological aspects behind this relation are elaborated in a recent review article~\cite{wen2019}.

The possibility of spinon and other fractionalized excitations triggered major interest in theories of quantum spin liquids. Experimental work inspired by their predictions initially focused on the search for materials that evade magnetic order as best candidates for real-world hosts of spinon excitations, although {\cred some of the recent results suggest that even long-range-ordered magnets may show non-trivial spectral features~\cite{piazza2015,banerjee2017}} potentially related to fractionalization. The question of whether these complex excitations are truly fractionalized is, however, much more subtle~\cite{winter2017}, and we shall see this below.

Different types of spin liquids exist under the umbrella of RVB physics. In triangular systems, short-ranged valence bonds like those shown in Fig.~\ref{fig:triangular}a give rise to a gapped $Z_2$ quantum spin liquid~\cite{moessner2001,moessner2002} characterized by a topological order~\cite{ioselevich2002} with two types of excitations. Those at higher energies are spinons due to the breaking of valence bonds, whereas low-energy excitations are \textit{visons} related to the mutual arrangement of the valence bonds and topological order therein~\cite{ioselevich2002,ivanov2004}. Among many peculiar properties of these excitations, we note their potential usage in topological quantum computing~\cite{ioffe2002}.

When valence bonds occur between all atoms in the crystal, including distant neighbors, gapless spin liquid  characterized by purely spinon excitations is formed~\cite{anderson1987}. Spinons develop a Fermi surface and interact with the U(1) gauge field~\cite{baskaran1988}. The main lure of this U(1) or, colloquially, ``spinon-metal'' state is associated with high-temperature superconductivity, because fractionalization separates spin and charge and may allow charge propagate unhindered by magnetic effects. A comparatively recent review of this topic can be found in Ref.~\onlinecite{lee2006}.

Experimentally, both gapless and gapped quantum spin liquids are characterized by a continuum of spinon excitations conveniently probed by inelastic neutron scattering. Thermodynamic and transport measurements offer additional diagnostic tools. The ``short-ranged'' RVB states are gapped and should give rise to an exponential behavior of the magnetic susceptibility and specific heat. In contrast, the gapless U(1) quantum spin liquid in a triangular system is identified by the sub-linear $T^{\frac23}$ power-law behavior of the specific heat~\cite{motrunich2005}, although linear or even exponential (gapped) behavior may occur too when spinons interact, eventually reducing the gauge symmetry to $Z_2$~\cite{lee2007,galitski2007}. 

Predictions for the magnetic susceptibility~\cite{nave2007a} as well as optical~\cite{ng2007} and thermal~\cite{nave2007b} conductivities of spin liquids with spinon Fermi surfaces are available too. However, one has to be aware that by no means these predictions exhaust all possible experimental responses, nor the family of RVB states embraces all possible instances of quantum spin liquids. Many other types of spin-liquid phases have been envisaged by theory~\cite{savary2017,zhou2017}, and their experimental identification should be taken pragmatically. Any unconventional behavior, {\cred such as linear or sublinear evolution of the magnetic specific heat}, coupled with the absence of magnetic order and presence of an excitation continuum may be a strong, if not compelling signature of the spin-liquid physics. 

\subsection{Spin Hamiltonians}
\label{sec:hamiltonians}
Further evidence for the spin-liquid behavior may come from the fact that material in question lies close to the interaction regime where theory predicts breakdown of magnetic order with the formation of a spin liquid. For the purpose of the current review, we restrict ourselves to spin models with pair-wise interactions typical for insulating systems. Materials like organic charge-transfer salts approach or even undergo metal-insulator transitions and require more complex spin models that also include multi-spin terms, most notably ring exchange, on top of the pair-wise interactions. These additional terms play crucial role in stabilizing quantum spin liquids~\cite{powell2011}.

Anisotropic spin Hamiltonian of triangular antiferromagnets comprises several terms,
\begin{equation}
 \mathcal H= \sum_m\left[\mathcal H_m^{\rm XXZ}+\mathcal H_m^{\pm\pm}+\mathcal H_m^{z\pm}\right],
\label{eq:ham}\end{equation}
where $m=1$ describes nearest-neighbor interactions, $m=2$ describes second-neighbor interactions, etc. The first term of Eq.~\eqref{eq:ham} stands for the XXZ Hamiltonian~\footnote{Same interaction terms are often written as $J_m^zS_i^zS_j^z$ and $J_m^{\pm}(S_i^+S_j^- + S_i^-S_j^+)$~\cite{gingras2014}, where in the isotropic case ($\Delta=1)$ $J_m^{\pm}$ is twice smaller than $J_m^z$. This difference should be taken into account when comparing exchange parameters reported in different publications.},
\begin{equation}
 \mathcal H_m^{\rm XXZ}=J_m\sum_{\langle ij\rangle}(S_i^xS_j^x + S_i^yS_j^y+\Delta S_i^zS_j^z), 
\label{eq:delta}\end{equation}
and $\Delta$ is the extent of the XXZ anisotropy in the notation of Refs.~\onlinecite{zhu2018,maksimov2019}.

The second term describes diagonal components of the exchange beyond the XXZ model,
\begin{align}
 \mathcal H_m^{\pm\pm}=\sum_{\langle ij\rangle} 2J_m^{\pm\pm}[(S_i^x &S_j^x-S_i^yS_j^y)\cos\varphi_{\alpha}- \notag\\[-10pt]
 &-(S_i^xS_j^y+S_i^yS_j^x)\sin\varphi_{\alpha}],
\label{eq:pmpm}\end{align}
whereas the third term stands for the off-diagonal anisotropy,
\begin{align}
 \mathcal H_m^{z\pm}=\sum_{\langle ij\rangle} J_m^{z\pm}[(S_i^y &S_j^z+S_i^zS_j^y)]\cos\varphi_{\alpha}- \notag\\[-10pt]
 &-(S_i^xS_j^z+S_i^zS_j^x)\sin\varphi_{\alpha}].
\label{eq:zpm}\end{align}
Here, $\varphi_{\alpha}={0,\pm2\pi/3}$ is the bond-dependent pre-factor {\cred that reflects the rotation of the local coordinate frame under three-fold symmetry of the lattice.} 

The $\Delta=1$ and $J_m^{\pm\pm}=J_m^{z\pm}=0$ regime implies isotropic (Heisenberg) interactions. In systems of our interest, departures from this regime can be caused by the easy-plane anisotropy ($\Delta<1$), anisotropy in the $xy$ plane ($J_m^{\pm\pm}\!\neq\! 0$), or the off-diagonal anisotropy \mbox{($J_m^{z\pm}\neq 0$)}.

\subsection{Survey of magnetic ground states}
\label{sec:gs}
Contrary to Anderson's conjecture, Heisenberg antiferromagnets with purely nearest-neighbor interactions feature the long-range-ordered $120^{\circ}$ ground state~\cite{huse1988,jolicoeur1989} that also survives (sometimes with weak modifications) in the presence of easy-axis ($\Delta>1$) or easy-plane ($\Delta<1$) anisotropies~\cite{kleine1992,yamamoto2014,sellmann2015}. Only in the pure Ising limit will a partially disordered state occur~\cite{wannier1950}. This state is, however, not a quantum spin liquid, because Ising spins are classical. Large residual entropy indicative of a classical spin liquid is found indeed~\cite{wannier1950}.

Quantum spin liquid can be stabilized by interactions beyond nearest neighbors. The $J_1-J_2$ model of Heisenberg spins on the triangular lattice received ample attention and was shown to host a spin-liquid phase at $J_2/J_1\simeq 0.07-0.15$~\cite{zhu2015,pli2015,saadatmand2015,hu2015,saadatmand2016}, although the nature of this phase remains vividly debated, with both gapless~\cite{kaneko2014,iqbal2016,gong2019,hu2019} and gapped~\cite{zhu2015,bauer2017} scenarios being likely proposals. At higher $J_2/J_1$, a collinear stripe order (Fig.~\ref{fig:triangular}b) becomes stable~\cite{jolicoeur1990,chubukov1992,lecheminant1995}. 

Exchange anisotropy offers another route to the spin-liquid state(s). The effect of the $J_1^{\pm\pm}$ and $J_1^{z\pm}$ terms {\cred of Eqs.~\eqref{eq:pmpm} and~\eqref{eq:zpm}} largely resembles that of $J_2$, because they stabilize stripe order too~\cite{li2016b,luo2017}. A region of the quantum spin liquid phase separates these stripe states from the $120^{\circ}$ order~\cite{zhu2018,iaconis2018} and may thus appear even in the absence of interactions beyond nearest neighbors (Fig.~\ref{fig:triangular}c). Interestingly, this quantum spin liquid of the anisotropic $J_1$-only model is connected to the corresponding regime of the isotropic $J_1-J_2$ model~\cite{zhu2018,iaconis2018}, indicating the same (conceivably, Dirac~\cite{iaconis2018,hu2019}) type of a quantum spin liquid anticipated in triangular antiferromagnets if either of the $J_2$, $J_1^{\pm\pm}$, or $J_1^{z\pm}$ are properly tuned -- a positive message for the experiment. Another (dual) spin-liquid region has been identified in the limit of large $J_1^{z\pm}$~\cite{maksimov2019} but may be harder to reach in real materials, {\cred where $J_1^{\pm\pm}$ and $J_z^{\pm}$ are usually smaller than $J_1$}. It is also worth noting that none of these putative spin-liquid phases is directly related to the RVB states discussed previously.

\section{Co-based materials}
Many triangular antiferromagnets were studied over the years, but only a handful of them show close relation to the aforementioned physics, while others entail different types of structural deformations. We first discuss Co-compounds that fulfill our criterion of geometrical perfection, do not yet enter the spin-liquid state, but can be described by model Hamiltonians in the vein of Eq.~\eqref{eq:ham} and reveal highly non-trivial excitations.

\subsection{Single-ion physics}
\label{sec:co-single}
Co$^{2+}$ proved convenient for studying spin-$\frac12$ antiferromagnets and especially triangular systems. In the octahedral environment, Co$^{2+}$ ($3d^7$) features $S=\frac32$, but its orbital moment remains unquenched. Spin-orbit coupling then splits this manifold and separates the lower-lying Kramers doublet that becomes predominantly occupied at low temperatures, acting as an effective \mbox{spin-$\frac12$}~\cite{abragam}. Orbital degeneracy is thus lifted by the spin-orbit coupling without lowering the symmetry, and a regular atomic arrangement with the strong geometrical frustration can be preserved, in stark contrast to other spin-$\frac12$ ions. For example, Cu$^{2+}$, Ti$^{3+}$, and V$^{4+}$ are all subject to strong Jahn-Teller effects that distort triangular frameworks or even lead to their defragmentation~\cite{pen1997,mcqueen2008}.

The typical splitting between the ground-state \mbox{($j_{\rm eff}=\frac12$)} Kramers doublet and lowest excited state is of the order of $10-20$\,meV~\cite{boca2004}, suggesting that at least below 50\,K the magnetism is purely spin-$\frac12$. In the trigonal and hexagonal symmetries typical for triangular antiferromagnets, Co$^{2+}$ acts as an anisotropic magnetic ion with different $g$-values and different exchange interactions for the in-plane and out-of-plane directions~\cite{abragam}. The extent of this XXZ exchange anisotropy is relatively weak, though (Table~\ref{table:co}). 

Many of the Co$^{2+}$-based triangular materials are non-frustrated, because closely spaced Co$^{2+}$ ions feature ferromagnetic interactions arising from nearly $90^{\circ}$ superexchange pathways, as in CoCl$_2$ and CoBr$_2$~\cite{wilkinson1959} or $\beta$-Co(OH)$_2$~\cite{hunt2016}. Robust antiferromagnetism is only possible in materials with large Co--Co separations that naturally eliminate any couplings beyond $J_1$, because second-neighbor Co--Co distances of at least 10\,\r A are prohibitively large for the superexchange. With negligible $J^{\pm\pm}$ and $J^{z\pm}$ terms, Co$^{2+}$ triangular antiferromagnets are doomed to remain in the $120^{\circ}$-ordered state, but can be used to probe its interesting dynamics.

\subsection{Hexagonal perovskites and magnetic excitations}
\label{sec:co-spectra}
\begin{figure}
\includegraphics{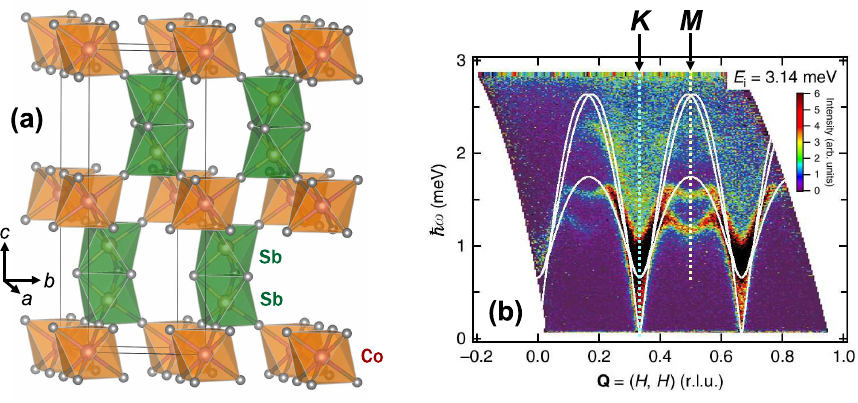}
\caption{\label{fig:co}
Excitations of the $120^{\circ}$ ordered state. (a) Crystal structure of 6H-Ba$_3$CoSb$_2$O$_9$ with the triangular layers of Co$^{2+}$ ions in the $ab$ plane. (b) Excitation spectrum probed by inelastic neutron scattering, with the white lines showing one-magnon dispersion from linear spin-wave theory~\cite{ito2017}. Note the excitation continuum that appears above 1\,meV at the $K$-point and above 1.8\,meV at the $M$-point, while at lower energies quasiparticle bands repelled by the continuum~\cite{verresen2019} are observed~\cite{ma2016,ito2017}. Panel (a) was prepared using the \texttt{VESTA} software~\cite{vesta}, {\cred and Ba atoms were omitted for clarity}. Panel (b) is reprinted from Ref.~\onlinecite{ito2017}, \copyright\ CC-BY-4.0.
}
\end{figure}

Best material prototypes of Co-based triangular antiferromagnets are found among hexagonal perovskites of the 6H-Ba$_3$CoX$_2$O$_9$ family with X = Sb~\cite{doi2004,shirata2012,zhou2012}, Nb~\cite{lee2014,yokota2014}, and Ta~\cite{ranjith2017,lee2017} (Fig.~\ref{fig:co}a). They share many similarities, but only the Sb compound was so far studied in detail thanks to the availability of large single crystals~\cite{prahabkaran2017}. It behaves as a typical easy-plane triangular antiferromagnet with the $120^{\circ}$ magnetic order in zero field~\cite{doi2004,zhou2012,ma2016} and anisotropic magnetization process revealing the $\frac13$-plateau at $10-15$\,T for in-plane fields, but no plateau when the field is applied perpendicular to the easy plane~\cite{susuki2013,quirion2015,sera2016}. High-field behavior of Ba$_3$CoSb$_2$O$_9$ was of significant interest, because the coplanar $V$-phase stabilized by quantum fluctuations~\cite{chubukov1991,griset2011} could be detected experimentally for the first time in a spin-$\frac12$ magnet~\cite{koutroulakis2015,liu2019}. However, the main draw of this material relates to its non-trivial magnetic excitations that were recently juxtaposed with theoretical predictions for triangular antiferromagnets.

Already the first theoretical studies of spin dynamics exposed salient deviations from non-interacting magnon scenario of linear spin-wave theory. Not only the spin waves are renormalized, sometimes changing their shape to produce roton minima~\footnote{These roton minima strongly influence thermodynamic properties of triangular antiferromagnets. Their relation to rotons in superfluid $^4$He is discussed in Ref.~\onlinecite{zheng2006b}.} around the $M$-points~\cite{zheng2006a,zheng2006b,ferrari2019}, but also the spectrum is washed out into a broad continuum at energies $\hbar\omega>J_1$~\cite{mezio2011,ghioldi2015,ghioldi2018,ferrari2019}. These findings were initially interpreted as signatures of spinon excitations -- an idea inspired by the work on square-lattice Heisenberg antiferromagnets. 

Spin-$\frac12$ antiferromagnets, especially in 2D, show reduced ordered moments indicative of spin fluctuations in the ordered state. This fluctuating component can be represented by an RVB state and held responsible for various spectral features~\cite{ho2001}, including the broadening of spectral lines at high energies interpreted as a spinon continuum~\cite{christensen2007,piazza2015}. While a similar physics could be envisaged in the triangular case~\cite{zheng2006b}, and spinons may indeed account for a large part of the calculated spectral weight~\cite{ghioldi2015}, non-linear spin-wave theory offered an alternative explanation in terms of interacting magnons. The roton minima can be reproduced by including the $1/S$ corrections~\cite{starykh2006,chernyshev2009} that are crucial in this case, given the nearly 60\,\% reduction of the ordered moment with respect to its classical value~\cite{capriotti1999,white2007}. Moreover, non-collinear spin arrangement facilitates magnon decays~\cite{chernyshev2006,zhitomirsky2013} that eventually account for the continuum at $\hbar\omega>J_1$~\cite{mourigal2013}. 

\begin{table}
\caption{\label{table:co}
Microscopic parameters of Co-based triangular antiferromagnets. All N\'eel temperatures $T_N$ and exchange constants $J$ are given in Kelvin. Two parameter sets for Ba$_3$CoSb$_2$O$_9$ are obtained from fits to the magnetization data~\cite{susuki2013} and to the inelastic neutron scattering data in the $\frac13$-plateau phase~\cite{kamiya2018}, respectively.
}
\begin{tabular}{c@{\hspace{0.5cm}}cccccc@{\hspace{0.5cm}}c}
\hline
 & $J_1$ & $\Delta$ & $T_N$ & $T_N/J^z$ & $g_{\|}$ & $g_{\perp}$ & Ref. \\
\hline 
Ba$_3$CoSb$_2$O$_9$ & 19.5 & 0.95 & 3.8 & 0.21 & 3.87 & 3.84 & \cite{susuki2013}\\
                    & 20.3 & 0.86 & 3.8 & 0.22 &      & 3.95 & \cite{kamiya2018}\smallskip\\
Ba$_2$La$_2$CoTe$_2$O$_{12}$ & 22 & & 3.8 & 0.17 & 3.5 & 4.5 & \cite{kojima2018}\smallskip\\
Ba$_8$CoNb$_6$O$_{24}$ & 1.7 & 1.0 & 0.1? & -- & \multicolumn{2}{l}{\quad 3.84} & \cite{rawl2017b} \\
\hline
\end{tabular}
\end{table}
Two aforementioned scenarios -- exotic fractionalized excitations vs. interacting magnons -- are in fact encountered in several materials of current interest, such as the Kitaev candidate $\alpha$-RuCl$_3$~\cite{banerjee2017,winter2017}. On the experimental side this leads to a large deal of ambiguity, because even the observation of an excitation continuum, the main fingerprint and ultimate signature of the spin-liquid physics, appears to be inconclusive when it comes to the question of whether spinons occur or magnons decay. 

Experiments on Ba$_3$CoSb$_2$O$_9$ confirmed that only at low energies, $\hbar\omega=J_1\leq 1.6$\,meV, do the excitations resemble magnons (Fig.~\ref{fig:co}b). At higher energies, a broad continuum is observed~\cite{zhou2012,ma2016,ito2017}. Magnons are strongly renormalized and damped~\cite{ma2016}, in agreement with non-linear spin-wave calculations. As for the continuum part of the spectrum, the bulk of the spectral weight is observed below 4\,meV~\cite{ito2017} corresponding to $\hbar\omega<2.5J_1$ in fair agreement with both spinon~\cite{ghioldi2015} and magnon~\cite{mourigal2013} scenarios. The aspect missing in both is the double-band structure around the $M$-point at $1.3-1.6$\,meV, (Fig.~\ref{fig:co}b) the energy range where excitation continuum already appears in other parts of the Brillouin zone~\cite{ito2017}. Recent theory work ascribed this feature, avoided quasiparticle decay, to an interaction between the one-magnon band and continuum~\cite{verresen2019}. Instead of smearing out the former, strong interaction separates the two. The continuum shifts to higher energies, while the ``one-magnon'' (quasiparticle) band is pushed down in agreement with the experimental observations. 

This strong-interaction scenario relies on the presence of an excitation continuum, while making no assumptions regarding its magnon or spinon origin. Indeed, whereas magnons are a good starting point for understanding all spectral features of large-spin triangular antiferromagnets~\cite{kim2019}, the spin-$\frac12$ case of Ba$_3$CoSb$_2$O$_9$ is more involved. First, vestiges of the continuum are seen up to much higher energies ($\hbar\omega\simeq 6J_1$) than any theory would predict. Second, exact parametrization obtained by inelastic neutron scattering in the $\frac13$-plateau phase of Ba$_3$CoSb$_2$O$_9$ does not allow an adequate quantitative description of the zero-field spectra even on the level of non-linear spin-wave theory~\cite{kamiya2018}. This leaves the problem of calculating spectral properties of $120^{\circ}$-ordered quantum antiferromagnet largely open, and the magnon-spinon dichotomy unresolved. 

\subsection{New materials}
Zero-field $120^{\circ}$ magnetic order generic to the 6H-Ba$_3$CoX$_2$O$_9$ materials appears to be one obstacle in the realization of a spin-liquid phase. This motivated several attempts to suppress magnetic order by increasing the distance between the Co$^{2+}$ layers. 

\begin{figure}
\includegraphics{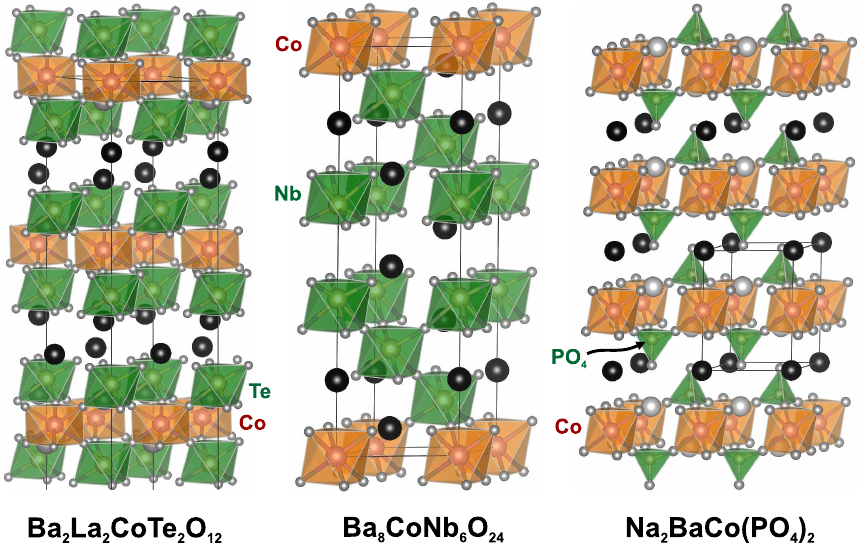}
\caption{\label{fig:co-structures}
Crystal structures of triangular Co-based antiferromagnets. The $c$ axis is along the vertical direction, whereas triangular layers are in the $ab$ plane. \texttt{VESTA} software~\cite{vesta} was used for crystal structure visualization.
}
\end{figure}

From the structural standpoint, triangular layers of Co$^{2+}$ are separated by perovskite-type slabs with non-magnetic ions. The thickness of these slabs can be increased (in theory, arbitrarily) leading to compounds like Ba$_2$La$_2$CoX$_2$O$_{12}$ (X = Te, W) with the interlayer Co--Co distance of about 9.8\,\r A and, eventually, to 
 Ba$_8$CoNb$_6$O$_{24}$, where the triangular planes of Co$^{2+}$ are as far as 19\,\r A apart (Fig.~\ref{fig:co-structures}). With the exception of the latter, these compounds still develop long-range magnetic order~\cite{rawl2017a,kojima2018}. In fact, even the Mn-based (spin-$\frac52$) analog of Ba$_8$CoNb$_6$O$_{24}$ reveals the $120^{\circ}$ order~\cite{rawl2019}, whereas Ba$_8$CoNb$_6$O$_{24}$ itself also shows a characteristic peak in the spin-lattice relaxation rate at 0.1\,K reminiscent of the magnetic ordering transition, although the NQR line does not broaden below this temperature~\cite{cui2018}. These observations suggest that residual interlayer couplings (likely of dipolar nature) always remain in place and tend to induce magnetic order with a sizable $T_N/J=0.1-0.2$. 

Despite the increased interlayer spacing and reduced $T_N$, even the most 2D materials show the same dynamics as their less 2D analogs~\cite{rawl2017b}. They can be most naturally understood as systems approaching eventual long-range order, which is indeed anticipated in any $J_1$-only XXZ antiferromagnet. An important lesson from these materials is that not only the interlayer couplings but also $J_1$ can become very small (Table~\ref{table:co}), thus shifting the onset of spin-spin correlations and $T_N$ to very low temperatures. Recently reported materials like glaserite-type Na$_2$BaCo(PO$_4)_2$ lacking magnetic order down to 50\,mK~\cite{zhong2019} should also be scrutinized from this perspective and do not necessarily feature spin-liquid behavior of any kind. Co$^{2+}$ vanadates of the same glaserite family entail ferromagnetic $J_1$~\cite{moeller2012,nakayama2013,sanjeewa2019}, suggesting the presence of a ferromagnetic exchange component that in phosphates may nearly cancel the antiferromagnetic one leading to a material with a very weak $J_1$ and only a weak frustration.

\section{YbMgGaO$_4$}
The Co$^{2+}$ compounds give access to only a limited part of the general parameter space of triangular antiferromagnets and remain in the region of the $120^{\circ}$ magnetic order. Other phases should be probed by systems with larger second-neighbor couplings and/or with a more pronounced exchange anisotropy. Anisotropic exchange interactions with sizable off-diagonal terms occur {\cred between $4f$ ions~\cite{wolf1971},} for example in spin-ice compounds with the three-dimensional pyrochlore lattice~\cite{rau2019}. The work on $4f$ magnets in 2D remained scarce until 2015 when synthesis and investigation of YbMgGaO$_4$~\cite{li2015a} suggested the possibility of a spin-liquid state and spurred interest in $4f$-based triangular antiferromagnets. In contrast to the Co$^{2+}$ compounds, where XXZ interaction regime is known with all certainty, the $4f$ ions with their different electronic configurations and crystal-field ground states offer (at least potentially) a much broader diversity of microscopic scenarios. An inevitable drawback is that each material features many exchange parameters leaving the microscopic parametrization ambiguous and the physical scenario controversial.

Most of the $4f$-based triangular antiferromagnets reported to date utilize Yb$^{3+}$ as the magnetic ion. We shall discuss its behavior in greater detail using the best studied material YbMgGaO$_4$ as an example. A more detailed and technical overview of this material can be found in a recent progress report~\cite{li2019r}.

\begin{figure*}
\includegraphics{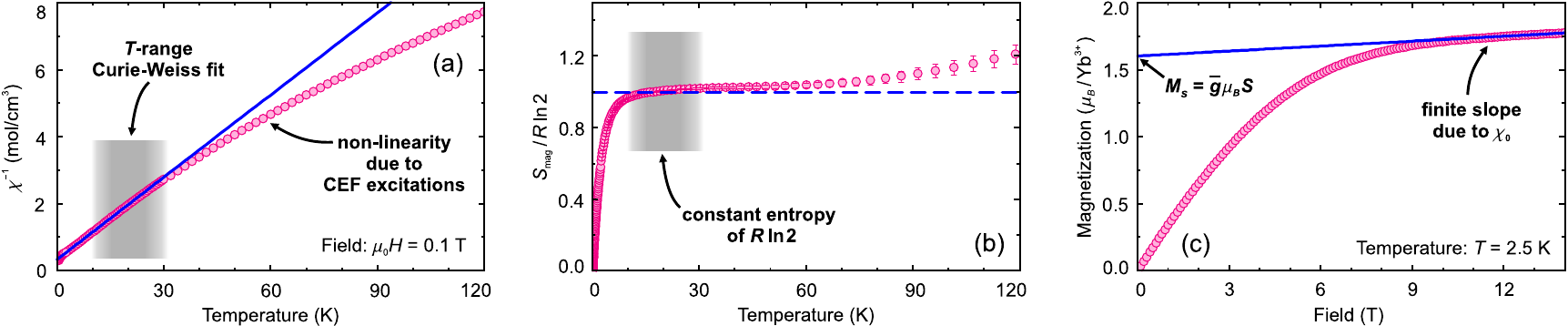}
\caption{\label{fig:yb-thermo}
Determination of the exchange parameters in YbMgGaO$_4$~\cite{li2015a}. The Curie-Weiss fitting of the inverse susceptibility, $\chi^{-1}$ (a), is performed in the temperature range of the $R\ln 2$ plateau in the magnetic entropy, $S_{\rm mag}$ (b), whereas the temperature-independent van Vleck contribution $\chi_0$ can be cross-checked by the slope of the magnetization isotherm, $M(H)$, above the saturation field (c). The saturated magnetization is $M_s=\bar g\mu_BS\simeq 1.6$\,$\mu_B$/Yb$^{3+}$ compatible with $S=\frac12$ and $\bar g\simeq 3.29$, the powder-averaged $g$-value determined from ESR~\cite{li2015b}.
}
\end{figure*}

\subsection{Nature of the Yb$^{3+}$ magnetism}
The central part of YbMgGaO$_4$ are Yb$^{3+}$ ions with their valence electrons residing in the $4f$ shell, which is subject to a strong spin-orbit coupling. While formally a $J=\frac72$ ion, Yb$^{3+}$ reveals an effective spin-$\frac12$ physics at low temperatures, because crystal electric fields (CEFs) split the $J=\frac72$ multiplet into four Kramers doublets, similar to the effect of spin-orbit coupling on Co$^{2+}$ (Sec.~\ref{sec:co-single}). Only the lowest CEF level is relevant to cooperative magnetism observed at low temperatures. This level is a Kramers doublet and can be directly mapped onto a spin-$\frac12$ problem. More precisely, one refers to the magnetic moment of Yb$^{3+}$ associated with this doublet as a pseudospin-$\frac12$, because it combines strongly intertwined spin and orbital moments. Its principal feature is magnetic anisotropy caused by the complex nature of the ground-state wavefunction and by the influence of the higher-lying CEF levels on the exchange. 

Another crucial feature of Yb$^{3+}$ is the strong localization of its $4f$ electrons. Despite this ultimate localization, magnetic interactions between the pseudospins are not of purely dipolar nature and involve orbital overlap. For example, in YbMgGaO$_4$ with the Yb--Yb distance of 3.85\,\r A, the dipolar interaction of 0.25\,K makes only 14\% of the total interaction $J_1\simeq 1.8$\,K determined experimentally from the Curie-Weiss temperature~\cite{li2015b}. This puts forward superexchange as the main mechanism of magnetic couplings in Yb-based triangular antiferromagnets and even allows a microscopic evaluation of magnetic interactions based on superexchange theory~\cite{rau2018}.

The superexchange is weak, though, so Yb$^{3+}$ oxides do not reveal their interesting cooperative magnetism unless cooled down to temperatures of the order of 1\,K. Many of these compounds were known since decades but traditionally described as simple paramagnets until measured at low enough temperatures. This naturally sets a question of the relevant temperature scale. Which temperature is low enough, and how to decide whether the absence of magnetic order down to a certain temperature is a signature of the spin-liquid behavior, or simply a feature of very weak magnetic interactions that are easily overridden by thermal fluctuations within the selected temperature range?

\subsection{Low-temperature behavior}
\label{sec:lowT}
YbMgGaO$_4$ serves as a good illustration of how this temperature scale can be determined experimentally. The strength of magnetic interactions is gauged by the Curie-Weiss temperature $\theta$, but the Curie-Weiss fit, $\chi=C/(T-\theta)+\chi_0$, must be done in a prudently chosen temperature range. At high temperatures (above $30-50$\,K in Yb$^{3+}$ oxides), $1/\chi$ deviates from the linear behavior due to CEF excitations (Fig.~\ref{fig:yb-thermo}a), while at low temperatures, typically below 10\,K, magnetic interactions come into play. Therefore, both upper and lower limits of the fit require a careful consideration. 

\begin{figure}
\includegraphics{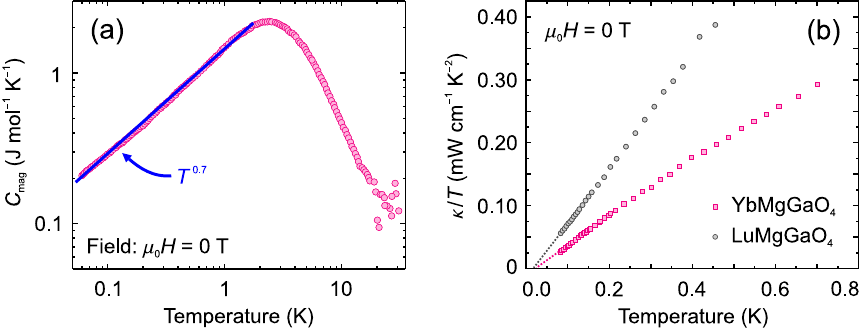}
\caption{\label{fig:yb-lowT}
Low-temperature properties of YbMgGaO$_4$. (a) Power-law scaling of the magnetic specific heat $C_{\rm mag}$ measured in zero field~\cite{li2015a}, and (b) thermal conductivity $\kappa$ suppressed with respect to the non-magnetic reference compound LuMgGaO$_4$~\cite{xu2016}. Dotted lines are extrapolations that highlight the absence of the linear contribution, which would be expected in a gapless quantum spin liquid.
}
\end{figure}

Magnetic entropy guides the choice. Between 10 and about 30\,K, it shows a plateau at $R\ln 2$ corresponding to the ground-state doublet (Fig.~\ref{fig:yb-thermo}b). The plateau implies that, on one hand, magnetic interactions have been overridden by thermal fluctuations and, on the other hand, the temperature is not high enough to trigger CEF excitations, although the very presence of the higher-lying CEF levels causes the non-zero van Vleck term $\chi_0$ in the susceptibility. Fitting three parameters ($\chi_0$, $\theta$, and the Curie constant $C$) in such a narrow temperature window certainly becomes ambiguous, but the estimate of $\chi_0$ can be cross-checked by measuring field dependence of the magnetization, because above saturation $M(H)$ still shows a small linear slope caused by the $\chi_0H$ contribution due to the CEF excitations (Fig.~\ref{fig:yb-thermo}c). As for the Curie constant $C$, it is proportional to the square of the $g$-factor that, in turn, can be independently determined from electron spin resonance (ESR). This leaves $\theta$ as the only independent fitting parameter and eventually leads to the estimates of $J_1\simeq 1.8$\,K and $\Delta\simeq 0.55$ from the inverse susceptibility measured for different field directions~\cite{li2015b}.

The interaction strength of about 2\,K also manifests itself in other physical quantities. Magnetic specific heat (Fig.~\ref{fig:yb-lowT}a) shows a maximum at 2.5\,K due to short-range order~\cite{li2015a}, whereas muon relaxation rate gradually increases around the same temperature, indicating the onset of spin-spin correlations~\cite{li2016a}. At temperatures well below 2\,K, one expects that spin-spin correlations dominate over thermal fluctuations, and experimental response of the magnetic ground state can be probed. Only via measurements in this low-temperature range can one assure that the given material hosts a spin-liquid state or at least bears relation to the spin-liquid physics.

In YbMgGaO$_4$, several experimental techniques probed the nature of the ground state. First, magnetic specific heat shows no anomalies down to at least 50\,mK~\cite{li2015a}, whereas muons reveal persistent spin dynamics~\cite{li2016a}: two necessary conditions of the spin-liquid behavior. Below 0.4\,K, magnetic specific heat follows the $C_m(T)\sim T^{\gamma}$ power law with $\gamma\simeq 0.7$~\cite{li2015a}, indicating gapless spin excitations (Fig.~\ref{fig:yb-lowT}). Moreover, the $\gamma$ value is compatible with $\frac23$ expected for the U(1) quantum spin liquid in triangular antiferromagnets (Sec.~\ref{sec:phenomenology}). 

\begin{figure}
\includegraphics{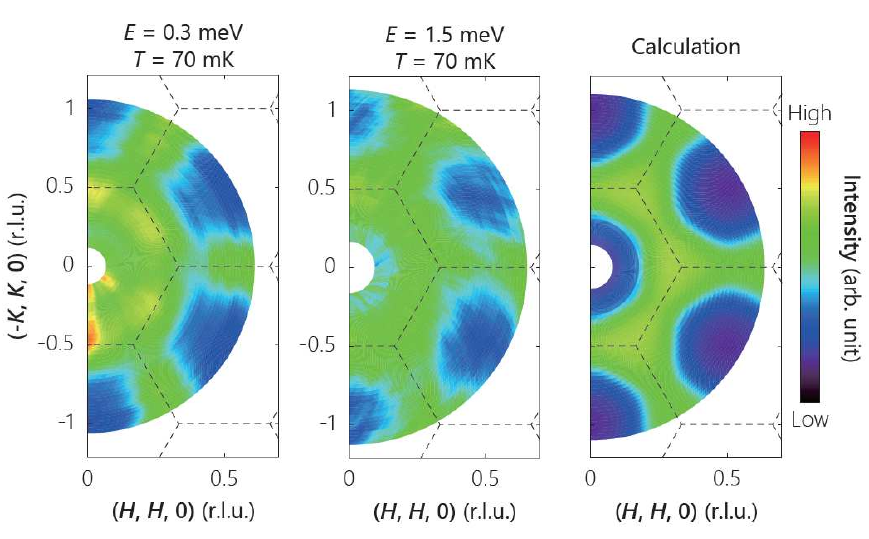}
\caption{\label{fig:yb-continuum}
Excitation continuum in YbMgGaO$_4$ at energies of 0.3\,meV ($\hbar\omega=2J_1$) and 1.5\,meV ($\hbar\omega=10J_1$)~\cite{shen2016}. Right panel shows the calculation based on the spinon-metal scenario. Reprinted from Ref.~\onlinecite{shen2016} with the permission, \copyright\ Macmillan Publishers Limited, 2016.
}
\end{figure}

Fractionalized (spinon) excitations of the U(1) state should also manifest themselves in the magnetic response and thermal transport. These experiments did not arrive at a consistent picture, though. YbMgGaO$_4$ shows no thermal transport due to spinons (Fig.~\ref{fig:yb-lowT}b). Moreover, its thermal conductivity is suppressed with respect to the non-magnetic Lu-based analog suggesting the absence of mobile spinons~\cite{xu2016}. ac-susceptibility measurements even revealed signatures of spin freezing around 100\,mK~\cite{ma2018}, but with only a small magnitude of the cusp and without any associated peak in the specific heat, although in spin glasses entropy change at the freezing point is usually detectable~\cite{schmidt2001}. Moreover, no splitting between field-cooled and zero-field-cooled curves measured in dc-field has been observed~\cite{li2019}. Magnetic response of YbMgGaO$_4$ is thus very different from that of a canonical spin glass -- perhaps not surprising, given persistent spin dynamics probed by muons even below the alleged freezing point~\cite{li2016a}. All these observations suggest that YbMgGaO$_4$ does show dynamic spins and hosts a spin liquid of some kind, whereas the peak in the ac-susceptibility may reflect only a minor frozen component.

\subsection{Spin dynamics}
\label{sec:dynamics}
Magnetic excitations of YbMgGaO$_4$ do not contribute to heat transport. Nevertheless, they do form a continuum that has been extensively studied by neutron scattering~\cite{shen2016,shen2018,paddison2017,li2017b,li2019}. It extends to about 2\,meV corresponding to energies as high as $\hbar\omega\simeq 13J_1$. With a very similar momentum dependence of the spectral weight at low (0.3\,meV) and high (1.5\,meV) energies~\cite{shen2016}, this spectrum (Fig.~\ref{fig:yb-continuum}) is drastically different from the typical response of $120^{\circ}$-ordered Co-based antiferromagnets (Sec.~\ref{sec:co-spectra}), where the spectral weight is restricted to much lower energies and shows different momentum dependence at different energies, even inside the continuum.

\begin{figure*}
\includegraphics{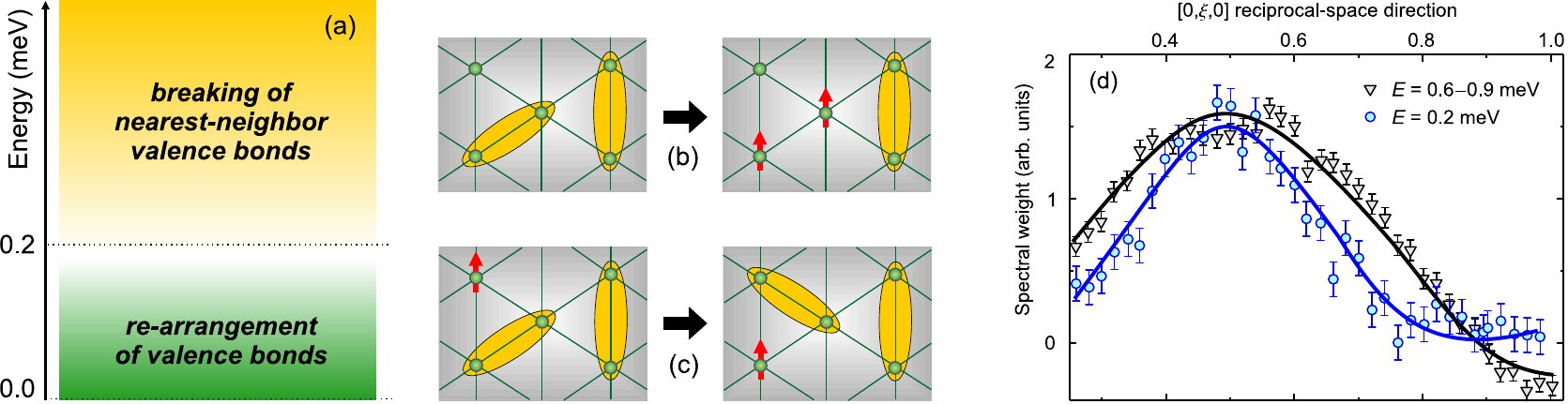}
\caption{\label{fig:yb-excitations}
Valence-bond model of the YbMgGaO$_4$ excitations~\cite{li2017b,li2019}. (a) Cartoon representation of the excitation continuum comprising two types of processes: (b) breaking of nearest-neighbor valence bonds, and (c) re-arrangement of valence bonds and orphan spins, where only nearest-neighbor valence bonds are shown for the sake of clarity, but valence bonds up to third neighbors are included in the actual model. (d) Momentum dependence of the spectral weight at two energies in different parts of the continuum~\cite{li2019}.
}
\end{figure*}

In YbMgGaO$_4$, the spectral weight accumulates at the zone boundary and appears to be nearly absent at the zone center, similar to the kagome spin liquid pinpointed experimentally in herbertsmithite~\cite{han2012}. Gapless spinon excitations of the U(1) quantum spin liquid can explain this distribution of the spectral weight down to at least 0.3\,meV~\cite{shen2016}. This scenario was further supported by a V-shaped band splitting in the applied magnetic field~\cite{shen2018} predicted theoretically for non-interacting spinons~\cite{li2017c}, although interactions between spinons change the scenario qualitatively~\cite{balents2019}. 

An alternative scenario was offered in Refs.~\onlinecite{li2017b,li2019} that interpret the same broad continuum as excitations out of a generic valence-bond state. Neutron scattering experiments down to 0.07\,meV~\cite{li2019}, about 4 times lower energy than in Ref.~\onlinecite{shen2016}, suggest a change in the width of the continuum with the threshold value around 0.2\,meV, which is comparable to the interaction strength $J_1$ (Fig.~\ref{fig:yb-excitations}). Above this energy, the excitations can be assigned to the breaking of nearest-neighbor valence bonds, while below 0.2\,meV the spectral weight is described by processes that involve a re-arrangement of valence bonds and orphan spins, with valence bonds up to third neighbors included in the model~\cite{li2019}. 

Interestingly, both pictures entail spin-$\frac12$ excitations albeit with a very different physics behind them. The spinon-metal scenario of Ref.~\onlinecite{shen2016} postulates fractionalized nature of the excitations without explicating their microscopic origin. It can be better traced in the valence-bond scenario, although here not all spin-$\frac12$ excitations are fractionalized. The continuum of high-energy excitations indicates that, upon breaking a valence bond, two unpaired spins can separate, similar to spinon excitations of an RVB state. These spinon-like excitations show a gap of about 0.2\,meV, which is on the order of $J_1$ and in agreement with theory (Sec.~\ref{sec:phenomenology}). On the other hand, lower-energy excitations caused by the re-arrangement of valence bonds fill this gap, extend to lowest energies, and stem from the presence of orphan spins in the ground state. These excitations are fractional but not fractionalized. Their presence -- or, more precisely, the inelastic nature of the re-arrangement process -- further indicates a departure from the simple RVB state, where interchanging of unpaired spins and valence bonds should cost no energy. The finite energy cost of this process may stem from a local non-equivalence of different lattice bonds that, in turn, has to be caused by structural inhomogeneities.

\subsection{Structural randomness}
\label{sec:randomness}
Whether or not YbMgGaO$_4$ is prone to structural inhomogeneities has been a matter of significant discussion. On one hand, this material was put forward as a geometrically perfect triangular antiferromagnet based on its robust trigonal symmetry that renders all nearest-neighbor Yb--Yb distances as well as relevant superexchange pathways equal~\cite{li2015a}. The absence of any detectable magnetic impurities~\cite{li2015b} and the lack of spin freezing (beyond the effects discussed in Sec.~\ref{sec:lowT}) would also imply that this material is less likely to suffer from inhomogeneities and disorder than other spin-liquid candidates, such as herbertsmithite~\cite{norman2016}. On the other hand, YbMgGaO$_4$ is still imperfect in the sense that Mg and Ga are randomly distributed in the non-magnetic slabs that separate the triangular planes of Yb$^{3+}$ (Fig.~\ref{fig:yb-structures}). Although such a mixture of non-magnetic atoms \textit{between} the magnetic places would normally have little effect on the magnetism \textit{within} these planes, the YbMgGaO$_4$ case appears to be different. 

\begin{figure*}
\includegraphics{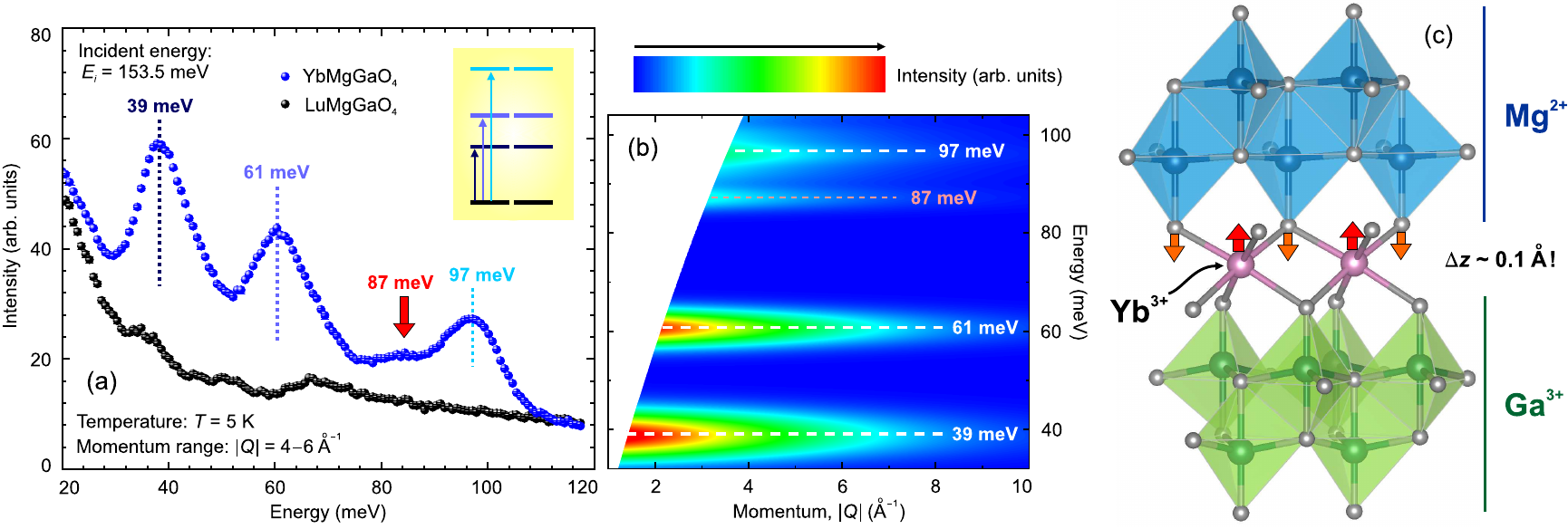}
\caption{\label{fig:yb-cef}
CEF excitations of YbMgGaO$_4$~\cite{li2017a}. (a) Inelastic neutron scattering reveals an additional feature around 87\,meV incompatible with four Kramers doublets of Yb$^{3+}$ (momentum dependence and the data for LuMgGaO$_4$, the non-magnetic reference compound, exclude possible phonon origin of this excitation). (b) Simulation of the experimental spectrum combined from several local configurations obtained with different distributions of Mg$^{2+}$ and Ga$^{3+}$ around Yb$^{3+}$; note that the 87\,meV feature appears as the side peak of the highest CEF excitation. (c) Sample local configuration with opposite displacements of Yb$^{3+}$ and O$^{2-}$ caused by the uneven charge distribution.
}
\end{figure*}
Unequal charges of Mg$^{2+}$ and Ga$^{3+}$ prove to be crucial. Depending on the local arrangement of these non-magnetic species, the Yb$^{3+}$ ions experience different CEFs. Inelastic neutron scattering reveals three CEF-related peaks, each of them being much broader than the instrumental resolution (Fig.~\ref{fig:yb-cef}a). Moreover, a shoulder around 87\,meV would indicate a ``fourth'' CEF excitation forbidden for Yb$^{3+}$, since only three Kramers doublets are available for excitations. Ref.~\onlinecite{li2017a} offered an atomistic interpretation of this strange CEF spectrum. The local arrangement of Mg$^{2+}$ and Ga$^{3+}$ creates an uneven charge distribution and causes local displacements of both Yb$^{3+}$ and surrounding oxygens (Fig.~\ref{fig:yb-cef}c). The magnitude of these displacements being as large as 0.1\,\r A implies that the CEF energies may change compared to the undistorted scenario. A superposition of several local configurations obtained within this approach leads to a decent description of the experimental spectrum (Fig.~\ref{fig:yb-cef}b).

Whereas CEF excitation energies do not determine magnetic interactions \textit{per se}, they influence the $g$-values that are spread over finite ranges $\Delta g_{\perp}/g_{\perp}\simeq 0.1$ and $\Delta g_{\|}/g_{\|}\simeq 0.3$, respectively. Moreover, local displacements change the Yb--O--Yb angles that may not affect relative values of the exchange parameters, but do change their absolute values. For example, from the superexchange theory of Ref.~\onlinecite{rau2018} one expects that the absolute value of $J_1$ varies by about 50\% throughout the crystal.

Other experiments support not only the presence of this structural randomness, but also its tangible effect on the magnetism. First, absent magnetic contribution to the thermal conductivity~\cite{xu2016} is naturally explained by the random exchange couplings that will cause localization of magnetic excitations regardless of their exact origin. Second, inelastic scattering in the fully polarized state above 7.5\,T shows an abnormally broad distribution of the spectral weight and hardly resembles spin wave of a ferromagnet~\cite{paddison2017,li2017a}. Third, the spin-liquid state of YbMgGaO$_4$ is remarkably insensitive to pressure~\cite{majumder2019}, suggesting that structural inhomogeneities may be instrumental in destabilizing magnetic order and facilitating spin dynamics. All these observations serve as the most direct evidence that structural randomness is central to the magnetism of YbMgGaO$_4$.

\subsection{Possible scenarios}
Several microscopic parameterizations reported for YbMgGaO$_4$ are summarized in Table~\ref{tab:yb}. Whereas all studies agree on the presence of easy-plane anisotropy ($\Delta<1$), more subtle (but crucial) details of the second-neighbor interactions and off-diagonal anisotropy remain controversial and illustrate challenges in the experimental determination of the exchange parameters for $4f$ magnetic ions. As many as four independent parameters have to be used for nearest-neighbor interactions, another four parameters can be envisaged for second-neighbor interactions, etc. 

\begin{table}
\caption{\label{tab:yb}
Exchange parameters for YbMgGaO$_4$ estimated by fitting different sets of the experimental data: i) Curie-Weiss temperatures and ESR linewidths~\cite{li2015b}; ii) inelastic neutron scattering in the fully polarized state and diffuse scattering in zero field modeled with~\cite{paddison2017} and without~\cite{li2018b} $J_2$; iii) inelastic neutron scattering and THz spectra~\cite{zhang2018}. The exchange parameters {\cred are introduced in Sec.~\ref{sec:hamiltonians} and} given in Kelvin with error bars where available.
}
\begin{tabular}{c@{\hspace{0.7cm}}c@{\hspace{0.7cm}}c@{\hspace{0.7cm}}c@{\hspace{0.7cm}}c}
\hline
           & I~\cite{li2015b} & II~\cite{paddison2017} & III~\cite{zhang2018} & IV~\cite{li2018b} \\\hline
$J_1$          & 1.8(2)         & 2.54(5)             & 1.98(7)          & 2.5  \\
$\Delta$       & 0.54(5)        & 0.58(2)             & 0.88(3)          & 0.76 \\
$J_1^{\pm\pm}/J_1$ & 0.09(1)    & 0.06                & 0.4(3)           & 0.26 \\
$J_1^{z\pm}/J_1$   & 0.02(4)    & 0                   & 0.6(6)           & 0.45 \\
$J_2/J_1$          & 0          & 0.22                & 0.18(7)          & 0    \\
\hline
\end{tabular}
\end{table}

\begin{figure}
\includegraphics{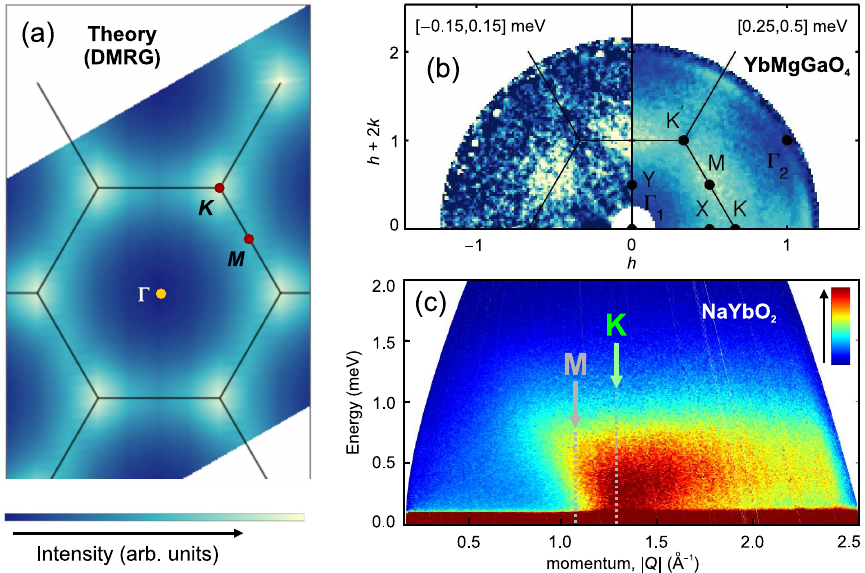}
\caption{\label{fig:yb-diffuse}
Neutron scattering from Yb-based triangular antiferromagnets. (a) Static structure factor obtained theoretically (DMRG) for the spin-liquid phase of the anisotropic spin Hamiltonian, Eq.~\eqref{eq:ham}~\cite{zhu2018}. (b) Momentum dependence of the spectral weight for YbMgGaO$_4$ (single crystal) with peaks at the $M$-points~\cite{paddison2017}. (c) Momentum dependence of the spectral weight for NaYbO$_2$ (powder sample) with the highest intensity at the $K$-point~\cite{ding2019}. Here, $K$ stands for the zone corner and $M$ for the midpoint of the zone edge. Panel (a) is reprinted with permission from Ref.~\onlinecite{zhu2018}, \copyright\  American Physical Society, 2018. 
}
\end{figure}

By disregarding $J_2$, the authors of Ref.~\onlinecite{li2015b} used Curie-Weiss temperatures and electron-spin-resonance (ESR) linewidths to determine all four components of the nearest-neighbor exchange tensor. This leads to a sizable XXZ anisotropy with the relatively weak but non-negligible additional terms $J_1^{\pm\pm}$ and $J_1^{z\pm}$ (Table~\ref{tab:yb}, set I). Fits to the inelastic neutron data~\cite{paddison2017} suggest a roughly similar interaction regime~\footnote{Note that the data of Ref.~\onlinecite{paddison2017} were rather insensitive to the $J_1^{\pm\pm}$ and $J_1^{z\pm}$ terms, so the values of Ref.~\onlinecite{li2015b} were employed there without further refinement.}, but with a sizable second-neighbor coupling $J_2/J_1\simeq 0.2$ that is assumed to be isotropic (set II). The primary reason for including $J_2$ is the peak of the neutron spectral weight at the $M$-points (Fig.~\ref{fig:yb-diffuse}b), as typical for the stripe phase at $J_2/J_1>0.15$. Alternatively, this stripe phase can be stabilized by the $J_1^{\pm\pm}$ and $J_1^{z\pm}$ terms (set IV) that describe the neutron data even in the absence of $J_2$~\cite{li2018b}. On the other hand, it was argued that these terms induce a large magnon gap~\cite{zhu2017} that is incompatible with the apparent gapless behavior~\cite{li2015a}. 

The accuracy of the neutron-based parametrization was improved by adding THz data that probe excitations at the zone center, as opposed to neutron scattering that is more sensitive to the zone boundary. This combined approach~\cite{zhang2018} hardly improves estimates of the $J_1^{\pm\pm}$ and $J_1^{z\pm}$ terms, but lends additional confidence in the finite-$J_2$ scenario (set III). The estimated value of $J_2=0.3-0.4$\,K largely exceeds the dipolar coupling of 0.07\,K and serves as the first experimental evidence of the long-range superexchange in Yb$^{3+}$ magnets.

With $J_2/J_1\simeq 0.2$, YbMgGaO$_4$ may not be far from the quantum spin liquid region of $J_1-J_2$ triangular antiferromagnets (Sec.~\ref{sec:gs}). An optimistic scenario would deem YbMgGaO$_4$ the first material prototype of this quantum spin liquid, but several experimental observations speak against such an interpretation. First, low-energy spectral weight peaks at the $M$-point~\cite{paddison2017}, while theory expects the peak at $K$~\cite{zhu2018} (Fig.~\ref{fig:yb-diffuse}). Second, absent magnetic contribution to the thermal transport~\cite{xu2016} leaves little room for a genuine, macroscopically entangled quantum state. Third, the randomness effect explicated in Sec.~\ref{sec:randomness} raises serious doubts about interpreting YbMgGaO$_4$ within the framework of any model having uniform exchange parameters.

The peak of the spectral weight at the $M$-point led to an idea that the ground state of YbMgGaO$_4$ may be a melted stripe order, where stripes are stabilized by the sizable $J_2$, but their directions are random under a change in the sign of $J_1^{\pm\pm}$~\cite{zhu2017} or a variation of all exchange parameters throughout the crystal~\cite{parker2018}, the latter scenario being consistent with the predictions of the superexchange theory~\cite{rau2018}. Both types of disorder lead to a liquid-like classical phase, although it remains unclear whether excitations of this random stripe state~\cite{zhu2017} may be responsible for the peculiar spin dynamics observed experimentally (Sec.~\ref{sec:dynamics}).

In fact, all ordered phases of triangular antiferromagnets are rather unstable toward randomness effects. Models with random distribution of nearest-neighbor exchange couplings were considered in the literature~\cite{watanabe2014,shimokawa2015} even before YbMgGaO$_4$ made a compelling experimental case for their relevance. Already weak randomness transforms the $120^{\circ}$ order into a glassy state~\cite{dey2019}, but the most interesting behavior is found in the limit of strong randomness and/or finite $J_2$, where a gapless spin-liquid-like phase distinct from either spin glass or valence-bond glass appears~\cite{wu2019}. This phase can be represented as a valence-bond state with short-range spin singlets and a small fraction of unpaired (orphan) spins~\cite{kawamura2019}. Alternatively, by analyzing a valence-bond state with random bond strengths, one can show that it is intrinsically unstable toward nucleation of spin-$\frac12$ topological defects, orphan or unpaired spins~\cite{kimchi2018}.

The above scenario is remarkably similar to the phenomenological model developed for the interpretation of magnetic excitations in Ref.~\onlinecite{li2019}. Indeed, a suitable parameterization of the valence-bond state with orphan spins allows quantitative description of the low-temperature thermodynamics, including the peculiar $T^{\frac23}$ power law of the specific heat~\cite{kimchi2018} that was initially ascribed to the U(1) quantum spin liquid. Ref.~\onlinecite{kawamura2019} describes this new, randomness-induced phase as a ``many-body localized RVB state'', but it clearly deviates from the conventional RVB scenario, because gapped vison excitations are preempted by the gapless low-energy dynamics of orphan spins. An interesting question is whether the mixed state of orphan spins and valence bonds, while not being an RVB state in the original sense, still shows quantum entanglement. First numerical results suggest that this may be the case~\cite{wu2019}, and give certain hope that the randomness-induced spin-liquid-like phase is not yet another case of the spin-liquid mimicry, but on a longer run may disengage itself from the precautionary ``-like'' ending.

\section{Other $4f$ materials}
The complexity of YbMgGaO$_4$ arises from the intricate combination of magnetic frustration and structural randomness. The former is essential, and the latter unavoidable, but a crucial problem on the materials side is whether this structural randomness can be reduced to the level that it does not change the physics qualitatively, leaving room for the ``genuine'' behavior of regular triangular antiferromagnets described in Sec.~\ref{sec:gs}.

\begin{figure}
\includegraphics{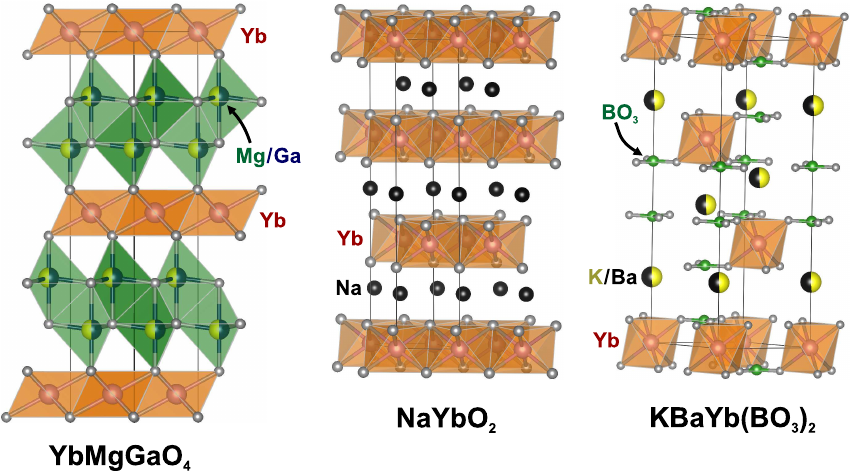}
\caption{\label{fig:yb-structures}
Crystal structures of Yb-based triangular antiferromagnets: YbMgGaO$_4$, NaYbO$_2$ as representative of the Yb$^{3+}$ delafossites, and KBaYb(BO$_3)_2$ as representative of the Yb$^{3+}$ borates. All structures are trigonal, with $c$ chosen as the vertical direction, whereas triangular layers are in the $ab$ plane. \texttt{VESTA} software~\cite{vesta} was used for crystal structure visualization.
}
\end{figure}
Close-packed layers of Yb$^{3+}$, the main building block of YbMgGaO$_4$, are common for many structure types, including the abundant family of delafossites, where the presence of only one sort of non-magnetic species should eliminate the randomness effect (Fig.~\ref{fig:yb-structures}). The absence of randomness was indeed confirmed in NaYbO$_2$~\cite{ding2019,ranjith2019,bordelon2019} by the observation of three resolution-limited peaks of the CEF excitations and the absence of extra features in high-energy inelastic neutron spectra~\cite{ding2019}. No significant randomness is expected in the isostructural materials NaYbS$_2$~\cite{baenitz2018}, NaYbSe$_2$~\cite{liu2018,ranjith2019b}, {\cred and CsYbSe$_2$~\cite{xing2019a}} as well. Interestingly, all these compounds reveal main signatures of the spin-liquid behavior, absent magnetic order~\cite{liu2018,ding2019,ranjith2019b,xing2019a} and persistent spin dynamics~\cite{baenitz2018,ding2019,sarkar2019} down to low temperatures. Moreover, in NaYbO$_2$ the spectral weight of the excitation continuum peaks at the $K$-point of the Brillouin zone~\cite{ding2019,bordelon2019} (Fig.~\ref{fig:yb-diffuse}c) in agreement with theoretical results for the quantum-spin-liquid phase of triangular antiferromagnets~\cite{zhu2018}. These promising observations suggest that the Yb$^{3+}$ delafossites may give experimental access to the spin-liquid phase in the absence of structural disorder.

So far little is known about the microscopic regime of the Yb$^{3+}$ delafossites. Average exchange couplings gauged by their Curie-Weiss temperatures and saturation fields are at least 2-3 times stronger than in YbMgGaO$_4$ (Table~\ref{tab:delafossites}), which shifts the relevant temperature scale toward higher temperatures, but simultaneously pushes the fully polarized phase above $14-16$\,T~\cite{ranjith2019b}, the feasibility limit of neutron-scattering experiments. The $g$-tensors determined by electron spin resonance are indicative of an easy-plane anisotropy (Table~\ref{tab:delafossites}) that may be even more pronounced than in the case of YbMgGaO$_4$. 

\begin{table}[!t]
\caption{\label{tab:delafossites}
$g$-values and exchange parameters of the Yb$^{3+}$ delafossites. The $g$-values are estimated by electron spin resonance~\cite{sichelschmidt2019}, whereas $J$ and $\Delta$ are calculated from Curie-Weiss temperatures obtained for two different field directions and represented as $\theta=\frac32 J$, where $J$ is the cumulative (nearest-neighbor and next-nearest-neighbor) coupling between spins pointing along this direction. Only the $J$ value is given for NaYbO$_2$ due to the lack of single crystals for this compound.}
\begin{tabular}{c@{\hspace{1.5cm}}c@{\hspace{0.5cm}}c@{\hspace{0.8cm}}c@{\hspace{0.5cm}}c@{\hspace{1cm}}c}
\hline
       & $g_{\|}$ & $g_{\perp}$ & $J$ & $\Delta$ & Ref. \\\hline
NaYbO$_2$ & 1.75 & 3.28 & 6.0 & -- & \onlinecite{ranjith2019} \\
NaYbS$_2$ & 0.57 & 3.19 & 9.0 & 0.13 & \onlinecite{baenitz2018} \\
NaYbSe$_2$ & 1.01 & 3.13 & 4.7 & 0.49 & \onlinecite{ranjith2019b} \\\hline
\end{tabular}
\end{table}
{\cred Another generic property of the Yb$^{3+}$ delafossites is their field-induced phase transition between the putative spin-liquid phase in zero field and one or several states with long-range magnetic order. In-plane magnetic fields trigger the transition already around 2\,T accompanied by the $\frac13$ magnetization plateau~\cite{ranjith2019b,xing2019a}, which is compatible with the up-up-down magnetic order confirmed experimentally for NaYbO$_2$~\cite{bordelon2019}. In contrast, out-of-plane fields cause magnetic order only above 8\,T with no plateau feature in the magnetization. This drastic difference in the transition fields reflects the sizable easy-plane anisotropy (Table~\ref{tab:delafossites}), whereas the presence (absence) of the magnetization plateau for the in-plane (out-of-plane) fields resembles the response of Co-based XXZ triangular antiferromagnets~\cite{susuki2013,quirion2015}, where interactions are restricted to nearest neighbors. Indeed, in the fully isotropic (Heisenberg) case, the up-up-down phase and associated $\frac13$-plateau occur only at $J_2/J_1<0.125$~\cite{ye2017a,ye2017b}, suggesting that in the Yb$^{3+}$ delafossites the $J_2/J_1$ ratio should be lower than in YbMgGaO$_4$, in agreement with the fact that the absolute value of $J_1$ increases.}

Other Yb-based triangular materials include the family of borate compounds ABaYb(BO$_3)_2$ (A = Na, K)~\cite{sanders2017,guo2019a,guo2019b}, where YbO$_6$ octahedra are not directly linked to each other, but connected via BO$_3$ triangles, with the nearest-neighbor Yb--Yb distance increasing to $5.3-5.4$\,\r A (Fig.~\ref{fig:yb-structures}). These materials may serve as interesting reference systems, where $J_2$ is effectively suppressed, giving way to the purely nearest-neighbor triangular model. The main question at this juncture is whether at least $J_1$ can be strong enough to produce any tangible magnetism. The Curie-Weiss temperatures well below 0.3\,K~\cite{guo2019a,guo2019b} suggest that an extremely cold environment would be needed to access frustrated behavior of these triangular antiferromagnets. With the shortest interlayer Yb--Yb distance ($5.3-5.4$\,\r A) approaching the intralayer one ($6.6-6.7$\,\r A), 2D nature of the magnetism also comes into question, whereas mixing of A$^+$ and Ba$^{2+}$ in the same crystallographic site creates random electric fields acting on Yb$^{3+}$, similar to the YbMgGaO$_4$ case (Sec.~\ref{sec:randomness}). 

Both delafossites~\cite{xing2020}, YbMgGaO$_4$-type~\cite{cevallos2018}, and mixed-cation borate~\cite{guo2019a,guo2019c} compounds exist for many if not all $4f$ ions. The properties of these materials remain to be explored, but a few literature cases suggest that at least Ce-based compounds will likely reveal an ultimate XY anisotropy reported for Ce$^{3+}$ in the trigonal CEF~\cite{banda2018}. {\cred Only a moderate easy-plane anisotropy has been observed in Er$^{3+}$ selenides, AErSe$_2$ (A = Na, K, Cs)~\cite{xing2019b,scheie2020}, but in this case the first crystal-field excitation lies below 1\,meV in the selenides~\cite{scheie2020} or around 2\,meV in the isostructural sulphide~\cite{gao2019} and is likely to impact the low-energy physics.}

Triangular magnetic layers are also featured by some of the non-chalcogenide $4f$ compounds. Most notably, CeCd$_3$P$_3$ shows signatures of quasi-2D magnetism and a long-range-ordering transition around 0.4\,K, although it undergoes a structural phase transition already at 127\,K~\cite{lee2019}. YbAl$_3$C$_3$ belongs to the same structural family and is known to develop singlet ground state~\cite{ochiai2007,kato2008} as a consequence of a similar structural phase transition~\cite{matsumura2008,khalyavin2013}. Metallicity of these compounds~\cite{lee2019} may be another important ingredient. Itinerant electrons will generally mediate long-range interactions, both within and between the triangular planes, thus rendering the mapping onto simple short-ranged spin models impossible or at least ambiguous.

Non-Kramers ions can be accommodated in the triangular geometry too, although they are known to produce Ising spins~\cite{nekvasil1990} that are quite interesting on their own right but leave little room for quantum fluctuations and spin-liquid behavior, at least in zero field. TmMgGaO$_4$ is an example of a triangular Ising antiferromagnet, where structural randomness plays a role as crucial as in the Yb$^{3+}$ analog, and the nature of magnetism remains controversial~\cite{shen2019,li2020}.

\section{Summary and Outlook}
Triangular antiferromagnets are no longer a land of static non-collinear magnetic structures of ever-increasing complexity~\cite{starykh2015}. Recent theory work charted stability regions of the quantum spin liquid (Sec.~\ref{sec:gs}) and identified unusual quantum features in the excitation spectra of the $120^{\circ}$-ordered state (Sec.~\ref{sec:co-spectra}). Structural randomness brought yet another dimension into this already rich phase diagram by making unexpected connections to valence-bond states that were historically proposed for triangular antiferromagnets and led to important conceptual developments but have been discarded as possible ground states of any realistic spin Hamiltonian.

On the experimental side, the research reviewed in this article leads us to reconsider the notion of the geometrically perfect spin-liquid material and the role that randomness or structural disorder play therein. The YbMgGaO$_4$ case shows that neither lattice symmetry nor complete site occupations are sufficient conditions of a ``perfect material'', because even subtle structural effects far away from the magnetic planes can strongly affect magnetic interactions and spin dynamics. This imposes very hard requirements for the material characterization, and renders probes like CEF excitations central to deciding whether a given material is ``perfect'' or not. On the more positive side, it also offers a convenient experimental knob for tuning the material toward a dynamically disordered state, potentially a spin liquid. 

The quest for ``structurally perfect'' spin-liquid materials was largely motivated by the common notion of imperfections being detrimental for the genuine spin-liquid state. A material with imperfections and without magnetic order was supposed to be a classical spin liquid that will freeze at low enough temperatures and show mundane spin dynamics. YbMgGaO$_4$ reveals this may not be the case, and even signatures of spin freezing (Sec.~\ref{sec:lowT}) do not preclude dynamic behavior for the majority of spins that, in fact, show highly unusual excitations. Taken together, these ideas suggest that structural imperfections should not always be avoided, but can be congenially used to facilitate spin dynamics. The general questions of which structural imperfections can be used in this way, and how to identify their influence on magnetic interactions, remain interesting avenues for future research in spin-liquid materials.

With the phase diagram of anisotropic $J_1-J_2$ triangular antiferromagnets fully charted and disorder-free materials like Yb$^{3+}$ delafossites already available, experimental access to the quantum spin liquid phase outlined in Sec.~\ref{sec:gs} becomes imminent. It makes, however, only one out of many interesting questions in the field. We outline a few others below.

First, with the exception of Ba$_3$CoSb$_2$O$_9$ field-induced behavior is largely unknown, and even theory studies of the anisotropic $J_1-J_2$ Hamiltonians in the applied field remain scarce~\cite{ye2017a,ye2017b}. NaYbO$_2$ shows a quite unusual transition between the spin-liquid phase in zero field and the collinear up-up-down phase in the applied field~\cite{ranjith2019,bordelon2019} (a similar transformation was claimed in randomness-influenced YbMgGaO$_4$ too~\cite{steinhardt2019}). Both phases may be quantum in nature, and their evolution is opposite to the typical scenario of Kitaev materials, where magnetic order is suppressed by the applied field giving way to a disordered state, possibly a spin liquid~\cite{winter2017b}. A quantum critical point separating the spin-liquid and up-up-down phases can be envisaged, with the Yb$^{3+}$ delafossites offering direct experimental access to it. 

Second, structural randomness could be tuned. YbMgGaO$_4$ lies in the limit of strong randomness, where valence bonds coexist with a significant fraction of orphan spins. Ramifications of this coexistence remain to be fully understood, but the most clear one is that abundant disorder renders magnetic excitations localized. An opposite limit of weak randomness, with valence bonds and only a minute fraction of orphan spins, may be, in contrast, the closest experimental realization of an RVB state. It can be conceived in Yb-based delafossites or other disorder-free triangular compounds that are suitably doped to induce weak randomness.

Third, all Yb-based triangular antiferromagnets studied so far are not magnetically ordered, which is even surprising because in other classes of materials finding an ordered state is by far easier than achieving a spin liquid. Tuning one of these materials toward an ordered state (or finding another Yb-based triangular compound with long-range magnetic order) may be an interesting endeavor. The primary goal here will be to probe magnetic excitations at different energy scales, keeping in mind the experience with Co-based compounds, where exact nature of the excitation continuum remains controversial. The proclivity of Yb$^{3+}$ and other $4f$ ions to the anisotropic exchange implies strong tendency to magnon breakdown with unusual spectral features that will benchmark theoretical studies of quantum spin-$\frac12$ triangular antiferromagnets.

Using theoretical map chart and appreciating the two-faced role of structural randomness will undoubtedly lead to new discoveries in the field of spin liquids and triangular antiferromagnets.

\acknowledgments
Intriguing and inspiring, triangular antiferromagnets and especially YbMgGaO$_4$ have triggered many fruitful collaborations and insightful discussions. Our special thanks goes to Sasha Chernyshev for his critical vision and comprehensive theoretical development of anisotropic triangular models, and to Sebastian Bachus and Yoshi Tokiwa for their impressive work in the \mbox{milli-K} lab. We are grateful to Devashibhai Adroja, Lei Ding, Franziska Gru{\ss}ler, Pascal Manuel, Astrid Schneidewind, and Qingming Zhang for common work on triangular antiferromagnets, to Martin Mourigal for his insightful and critical remarks, to Shiyan Li for providing thermal-conductivity data of YbMgGaO$_4$ for this review, and to Michael Baenitz for sharing unpublished results on the Yb$^{3+}$ delafossites. Last but not least, we thank German Science Foundation (DFG) for supporting our research on frustrated magnets via Project No. 107745057 (TRR80). AT further thanks Alexander von Humboldt Foundation for the financial support via the Sofja Kovalevskaya Award.

%

\end{document}